# Transit directions at global scale

Joris van der Geer
joriswu@gmail.com
13 Jan 2016

In this article I present a novel approach to integrated ground and air public transport journey planning, operating at continent scale. Flexible date search, prerequisite for long distance trips given their typical low and irregular service frequencies, is core functionality. The algorithm, best described as timed branch-and-bound (BBtime), is especially suited for irregular and poorly structured networks.
Almost all of the described functionality is implemented in a working prototype. Using ground transport only, the system is on par with Google Transit on random country-wide trips in the US.

## 1 Introduction

When travelling longer distances, flights often form part of the trip, with local bus and train serving the initial and final journey segment(s). Ideally, a journey planner can optimise whether – and if so, which – flight(s) are a good choice, taking initial and final segments into account in an integrated approach. This implicitly provides a 'search nearby airport' function by design. The typical lower frequency and/or less regular service of air and long-haul ground transport, the higher variation in minimum transfer times, as well as reservation requirement, warrant an approach that differs from commonly used algorithms. The overview given in [2] describes most algorithms. Raptor and CSA do no scale well to the high number of routes as is common in air traffic. Transit patterns, as described in [3]and [4], operate per departure time and hence become impractical at higher timetable variability and irregularity.

The system described here is unique in several aspects:
- single, unified algorithm integrating most modes of transport – albeit some with restrictions
- automatic, adaptive flexible date search
- fast precomputations ( relative to systems that do precompute)
- real-time support such as delays, seat / fare availability – with limitations

These aspects have been the main guideline to choose a suitable algorithm : timed branch-and- bound, henceforth referred to as BBTime. Secondary guideline is preferring to rely on implementation efficiency over algorithmic complexity, and avoiding approximations unless unavoidable.

## 2 Algorithm

### 2.1 Network

The network is represented as a time-dependent graph as in [1] with stations (bus stops, airports) as nodes, and direct connections as edges. Each edge represents a single transport service and carries a list of departure times, normalised to UTC. Schedule exceptions and irregularities are supported by not exploiting periodicity in this list, yet compressed based on repetition to reduce space by about a factor 5. Times are converted to UTC at data import, and to localtime in result reports.

## 2.2 Search

Goal of the search is to return the best trip given a departure time range. It is not formulated as an earliest arrival problem as the departure time is not considered given. The time range is not predetermined : the range is expanded such that the likelihood of finding better results outside the range are minimal. The criteria for *best* is a cost function that can be total trip time, or a biased value taking number of transfers, walk distances or taxi fares into account.

The core of the search algorithm consists of a global branch-and-bound optimisation with ancillary heuristic pruning. Initial bound is infinity. To make such approach feasible at larger scale, it is essential to provide bound opportunities in the branches close to the root. This is facilitated by precomputing shortest duration per connection and shortest transfer times per connection pair. End-to-end trip time and cost is evaluated at the tips of a branch, making bounds applied up to the last branch beneficial in terms of running time.

The first branch evaluates (Dep,Via) in trip (Dep,Via,Arr). Branch *n* evaluates ($Via_{n-1}$,...$Via_n$) in (Dep,Via1,...,$Via_{n-1}$,$Via_n$,Arr). At each branch, each station that has a direct connection with both preceding and succeeding stations qualifies as candidate for $Via_n$. End-to-end time for each complete trip reaching the destination not exceeding the current bound is assessed by searching for the combination of departure times at each segment such that the overall cost function is lowest. The cost function can to some extent take multiple criteria into account. The initial departure timespan is optionally determined at this stage for each assessed trip individually.

This optimisation is performed for a given number of transfers T, and hence *n* in $Via_n$ is constant. The trivial case of T=0 represents a direct connection between origin and destination and evaluates (Dep,Arr) directly.

The search proceeds iteratively over T=0..$T_{max}$, passing the existing bound at each iteration. Rationale for this order is that iterations for lower values of T are computationally less complex, more likely to result in lower times and thus lower the bound early. The bound mechanism guarantees a faster trip with more transfers will be found.

The search is sped up by precomputing complete trip fragments – henceforth named triplets – consisting of prospective connections. The search proper deals with triplets as if they were single segments, albeit decomposing into its constituents at appropriate evaluation steps.

Optionally, the graph is simplified by clustering geographically adjacent nodes together. The search will take the relative distances into account, yet in both precomputations and branch evaluations such cluster is treated as a single node.

## 2.3 Flexible dates

Earliest allowed departure time is given as query parameter. The allowed time span is dynamically adjusted such that a long transfer time or initial waiting time is avoided if an increase in span is likely to result in a departure time resulting in a better total trip cost.

## 2.4 Heuristics

The ratio $\frac{D_{route}}{D_{geo}}$ with $D_{route}$ the distance over the route and $D_{geo}$ the geographical distance, is taken as indirect likelihood for lower trip times and cost. Rationale is that shorter distances often involve lower time and fare, and the number of alternatives grows superlinear with higher ratio. Above a certain threshold *G*, candidate routes are ignored. For air transport, this threshold is typically set higher than for ground transport as the irregular fare structure warrant a more expensive yet broader search. The threshold is made to increase for shorter $D_{geo}$. In general, this threshold determines the broadness of the search space in geographical terms.

Time-limited priority branching is essentially providing a time limit on the branch-and-bound, with the branches visited in order of increasing distance and time. Thus, the most prospective branches are visited first.

## 2.5 Precomputation

Triplets between all station pairs are precomputed for each transfer T=[1,2] and stored in a sparse 2D matrix per T. Sparse allocation is used as the needed size would be quadratic in the number of station pairs. Due to the inherent clustered nature of the network, resulting storage size is more close to linear than quadratic. The direct connections for trivial case T=0 are stored in a similar matrix as well for quick access. Per station pair (Dep,Arr), all possible trips (Dep,Via,Arr) for T=1 and (Dep,$Via_1$,$Via_2$,Arr) for T=2 are evaluated. This is done by enumerating all possible combinations of single connections Dep-$Via_x$ and $Via_x$-Arr for each possible *x*. Each enumeration result is assigned a typical end-to-end time as follows :

```
for each departure time Tdep in random(now .. 2 weeks) do
  stop if Tdep > 64
  accu[Tdep] =  min_triptime(Dep-Via,Via-Arr) @ Tdep
done

avgtt = avg(accu)
accu[i] = accu[i] except abs(accu[i] – avgtt > threshold)

end2end_typ = avg(accu)
```

The min_triptime function searches for the best total time starting at Tdep up to a given timespan, honouring minimum transfer times appropriate for the mode of transport.

Effectively, a mini-trip planning is done with a statistically significant set of start times randomly chosen over one or more weeks, and averaged except outliers. Purpose of this approach is to take traffic days in account, anticipating that the search will avoid poor connections whenever possible.

A list of *N* triplets with lowest end-to-end time is stored per station pair (Dep,Arr) per T. N is 4 .. 16. The larger values for N are typically assigned to hub stations based on higher connectivity and/or reach. Rationale for N being well above 1 is manifold : account for transfer time differences due to traffic days or timetable variations, support alternatives regarding fare unavailability or delays, and most importantly, to compensate for the fact that the best connection between a pair of stations – a local optimum – does not guarantee such connection is the best choice for the entire trip – a global optimum –.

Calculations for T=2 on a 120K station network (Oceania) take 26 hours on a typical server and result in 1.5G triplets for 220 M pairs. For T=1 the values are 1 hour time, 400 M triplets and 46 M pairs.

# 3  Multimodal extension

Walking between nearby stations is a viable option, and is integrated into the core algorithm by pre-assessing walkable distances and adding these edges to the graph. Edges are assigned a duration based on a pedestrian road network planning and have no schedule. With maximum distances up to 1 or 2 Km, graph size doubles, which is acceptable. A similar figure is reported in [3]. The search can reject candidate trips based on a given maximum total walk distance, which is a query parameter.

Integrating unrestricted bicycle or car transport modes into the main graph would expand the graph considerably, and the derived structures prohibitively, as mentioned in [4]. We can restrict these modes by providing a subset of all possible routes based on them either being essential or deemed an especially beneficial option for the traveller. An example of the first case is a regional town without public transport. An example of the latter is an Uber/taxi connection between two airports.
Both scenarios are inferred from road network (Openstreetmap) and geolocation data. Mutual connections between a small number of suitable locations are generated based on station connectivity and combined direct connection reach.  Similar to walk edges, car and bicycle edges are integrated into the graph. In the first case, a node is added as well, representing the location.

True door-to-door planning is hard to integrate, as one would ideally need to evaluate a number of origin stations and a number of destination stations, adding the corresponding road segments from and to the ultimate origin and arrival locations into the equation. An approximation can be made as follows : define a geographical mesh with cell spacing A. Assign road-based connections from each mesh point towards one nearest suitable station in each direction. Dependent on the distance, this can be walk, bicycle or car. Skip this step for mesh points not nearby any road or not having any station in suitable distance. Add these connections to the graph. Adjust the search results by replacing the initial and final segment such that the mesh point is bypassed. This may imply a route planning on a road network at this stage for these segments individually. The spacing A is a measure of inaccuracy that needs to be balanced between desired accuracy and graph size.

# 4  Live fares and real-time support

Both aspects share in common data not being available at precomputation and graph build time.
The search graph contains a snapshot of timetable – and eventual fare – information at system preparation time. Precomputations derive from the former.

Live fares and by extension seat availability affect individual trips on given days. For ground transport, the affected origin and destination stations need to be given as well. The fares are imported dynamically and annotated as attribute on the corresponding departures / arrivals in the time list on the relevant edges in the graph. The search stage evaluating the times inspects these, incorporate fares into the cost function and seat availability by accepting/rejecting specific departures.

Delays and cancellations can be handled similarly by annotating adapted departure / arrival times and availability at relevant edge attributes. In this case, the search stage scanning candidate departures for each segment need to adjust the times and accepts/rejects departures based on the adjusted time.

To make this approach work, the search space needs to be defined broad enough to provide  sufficient alternatives. The parameter $N$ in § 2.5 and the threshold $G$ in § 2.4 define this space.

# 5 Refinements

A variant of a Dijkstra search without priorities is used to perform a profile search on each station in order to determine and diagnose global network connectivity. For each station, the total number of stations reachable within any number of transfers is determined, as well as the minimum number of transfers to do so. Schedules are not taken into account. Purpose is to assess whether each place is in principle reachable from each other place.

Secondary use is to construct an approximation to the minimum number of transfers needed to reach each station from each station. This would requires space quadratic in the number of stations. As this is prohibitive for larger networks, an approximation is made by storing the lowest value for any member of stations within the same two geographic mesh cells. The mesh size is chosen to balance storage size and accuracy and is stored in a sparse array. The search skips evaluating transfers below this minimum.

# 6 Experimental results

A prototype is implemented in C containing all functionality except door-to-door approximations and real-time delays. Full source code is available at Github https://github.com/joriswu/tripover.

All available timetable data for US transport agencies is loaded, together with timetable data for United and Qantas Airlines and all Skyteam and Oneworld member airlines. Long-distance ground transport data for Amtrak and Megabus were obtained separately.

The resulting network has 500 K stations and 20 M edges from 20 K routes. Edges have in total 5.65 G departures.

An Amazon Elastic Cloud r3.4xlarge instance is used, with 16 cores and 120 GB memory. Precompute stages use all cores, the search uses one core.

25 random queries were performed on a setup without airlines. The same queries were performed on Google Transit (GT) and Bing Transit (BT) for comparison. Appendix 1 lists the queries used. BT failed at 8 out of 15 queries and is left out of the comparisons. For 17 trip results, times differed less than 5% between GT and BBtime. For 1 query, BBtime had a 7% slower trip. For 2 queries, GT had slower trips by 29% and 20%. Two queries failed at GT. One query showed identical trips for both. Query times are on average 500 msec for GT and 900 msec for BBtime, albeit 24 out of 25 trips give identical results at 500 msec.

500 random queries were performed on an integrated ground and air transport setup. Only 1 query returned no result, due to an error in the data. Between 30K and 1M trips are evaluated per query. Query time is around 500 milliseconds, as determined by the search time limit. No indications of suboptimal results were found. As no other system can plan such trips, there is no way to verify the results other than manually inspect timetables. Appendix 2 shows 2 random results.

# 7 Discussion

Most design choices were driven by the intent to accommodate long-distance travel correctly and flexibly. Especially with integrated ground and air transport, the non-overtaking property as mentioned in [2] no longer holds. The less regular timetable structure makes an optimal route depend more significantly on departure time and day, and requires the search to have a broader scope and be less dependent on structure and periodicity.

Main drawback is high memory requirements for today's commodity server standards. As the system is not designed for distributed operation, there is no easy remedy other than seeking further optimisations.

Support for real-time aspects and fare availability is enabled in the design, yet its feasibility needs to be assessed. For this to work well, parameters need to be adjusted towards a broader search with correspondingly higher system requirements and more demanding search operation.

# 8  Conclusion

The system works well : query times are typically below 1 second, often below 500 msec, and results are on par with state-of-the art systems. In the US, the integration of air transport enables finding substantially faster trips. Any schedule and network irregularity can be accommodated without penalty. True multimodal operation and multicriteria search can be achieved to some extent.

The approach has shown to be remarkably robust, discovered when earlier prototype versions with impaired search functionality still gave good results despite the extent of the impairment suggesting otherwise.

Setting up a system from scratch, starting with original timetable data, takes a few days for continent-size setups and is fully automated. Hence, changes can propagate within a few days.

## 10  Data sources

Timetables:
- GTFS repository: transitfeeds.com
- Megabus: us.megabus.com
- Amtrak: Amtrak.com
- Oneworld: Oneworld.com
- Skyteam: Skyteam
- United Airlines: united.com

Geographical database: www.geonames.org/
Road network: Openstreetmap at download.geofabrik.de

## Appendix A

25 random trips were generated. Departure and arrival stations for the first 15 trips were chosen as a random element from the network station list. For the next 10 trips, random geographical coordinates within the USA bounding box were taken for both departure and arrival. The station nearest to each chosen point was taken.

Table 1 summarizes the properties of these random trips.
Table 2 shows the ground-only results for Google Transit (GT) and BBTime.
Table 3 shows the results for combined ground and air transport in comparison with Table 2.

The trip details are available at the project wiki :
USA test trips - ground only
USA test trips - ground plus air

| trip # | from | to | geo. dist | from coords | to coords |
|---|---|---|---|---|---|
| 1 | Broadway/w 219 st, Inwood, NY | Nova d/davie r – (mcfatter) Broward, FL | 1744 | 40.871966,-73.91304 | 26.086659,-80.232552 |
| 2 | 20 AV/128 ST, College Pt, NY | Yellowstone bl/austin st, Forest Hills, NY | 7 | 40.7816459,-73.8406140 | 40.7226640,-73.8518229 |
| 3 | Loughborough @ morganford eb, St Louis, MO | E 25 ST@E 6 AV, Miami, FL | 1697 | 38.5654959,-90.2762819 | 25.8446840,-80.2704080 |
| 4 | Sw 72 st@mentone st, Miami, FL | Rt-44 at robin rd, West Deptford, NJ | 1640 | 25.7049310,-80.2749989 | 39.8351780,-75.2190910 |
| 5 | Rt-70 at lexington blvd, Stockton, NJ | School of Education, Amherst, MA | 347 | 39.9309370,-75.0622999 | 42.3976209,-72.5279460 |
| 6 | Park ave & newington sb, Baltimore, MD | Lakeshore Dr at Elkwood Ave, Asheville, NC | 665 | 39.3138559,-76.6292849 | 35.6297760,-82.5748420 |
| 7 | Interbay Blvd @ Trask St, Hills City, FL | Copans r/banks r, Broward, FL | 292 | 27.8659739,-82.5224390 | 26.2549190,-80.1963200 |
| 8 | Steamboat Rd / Middle Neck Rd, Nassau, NY | Hanover & clement s/b, Baltimore, MD | 299 | 40.8085820,-73.7359579 | 39.2736789,-76.6154410 |
| 9 | W 88th Ave & Lipan St, Denver, CO | Eighth St & Folsom St, SF, CA | 1523 | 39.8563800,-105.0013929 | 37.7748140,-122.4099879 |
| 10 | Pleasant st + rockwood ave, Worchester, MA | Hwy. 116 & Woodworth Rd, Sonoma, CA | 4273 | 42.2806590,-71.8707899 | 38.3493760,-122.7627899 |
| 11 | Eliot av/metropolitan av, Flushing, NY | Michigan & 29th Street, Chicago, IL | 1152 | 40.7129810,-73.9047619 | 41.8417379,-87.6236240 |
| 12 | Hillview Rd & Eastwood Rd, Minneapolis, MN | Santa Anita / La Sierra, LA, CA | 2437 | 45.1115749,-93.2132869 | 34.1263840,-118.0313189 |
| 13 | Olive @ partridge sb, St Louis, MO | NW Macleay & Lomita Terrace, Portland, OR | 2758 | 38.6679679,-90.3196240 | 45.5263129,-122.7069880 |
| 14 | Church St. at Palisade Dr, Santa Maria, CA | Tweedy / San Juan, LA, CA | 232 | 34.9519820,-120.4133270 | 33.9438170,-118.2032250 |
| 15 | Abrams @ birch - n – fs, Dallas, TX | Rt-77 at cumberland ave, Bridgeton, NJ | 2056 | 32.9316539,-96.7334060 | 39.4444480,-75.2216399 |
| 16 | Lompoc-Surf Amtrak, CA | Quincy Amtrak, IL | 2638 | 34.6827,-120.6050 | 39.9571,-91.3685 |
| 17 | Maysville Amtrak, KY | Galveston, TX | 1452 | 38.6521,-83.7711 | 29.3066,-94.7968 |
| 18 | Branson / Hollister, MO | Vale, OR | 2185 | 36.6184,-93.2279 | 43.9824,-117.2482 |
| 19 | Sandstone, MN | Port Huron Amtrak, MI | 898 | 46.1243,-92.8856 | 42.9604,-82.4438 |

| 20 | Glasgow Amtrak, MT | Upton & nimmo, Virginia Beach, VA | 2790 | 48.1949,-106.6362 | 36.7591,-76.0057 |
| 21 | McGregor, MN | Alpine Amtrak, TX | 2015 | 46.6078,-93.3009 | 30.3573,-103.6615 |
| 22 | Sr 167 & S Grady Way, Rento, WA | Benton Station, CA | 1116 | 47.4713,-122.2177 | 37.8187,-118.4777 |
| 23 | Malta Amtrak, MT | Missoula, MT | 489 | 48.3605,-107.8722 | 46.8798,-114.0203 |
| 24 | Gillett, PA | Garden City Amtrak, KS | 2092 | 41.9255,-76.7974 | 37.9644,-100.8733 |
| 25 | Sault Ste. Marie, MI | Imperial Beach Bl & Seacoast Dr, Imperial Beach, CA | 3174 | 46.4635,-84.3717 | 32.5767,-117.1315 |

Table 1. properties for 25 random trips within the US

| Trip # | Google Transit | | BBTime | | +/- |
|---|---|---|---|---|---|
| 1 | 29h | TWBB[1] | 28h 11 min | BTTWBW | -/- |
| 2 | 57m | BTTW | 55m | BBW | -/- |
| 3 | 45h | BBBBBBW | 44h 50m | BWBBBBWBW | -/- |
| 4 | 29h | WTWTBTB | 31h | BBTTTB | +7% |
| 5 | 8h 24m | BBTWBB | 8h 4m | BBTWBW | -4% |
| 6 | n/a | | 13h 41m | BWTXBW | |
| 7 | 8h 8m | BBTBW | 8h | BBBBBW | -/- |
| 8 | 4h 30m | BTTBW | 4h 30m | BTTWBW | -/- |
| 9 | 35h | WBTTW | 34h 45m | WBTTB | -/- |
| 10 | 3d 22h | BTTTTBBBB | 3d 21h 23 m | BTTBBB | -/- |
| 11 | 20h | BTTBBB | 20h | WBTBB | -/- |
| 12 | 2d 16h | BBWBTBB | 2d 1h 26m | BBWBTWB | -29% |
| 13 | 2d 8h | BBTTBBW | 2d 6h | BWTTTWBW | -4% |
| 14 | 6h 39m | BBBWB | 6h 50m | WBTWBW | -/- |
| 15 | 1d 20h | WTWBBBBTBB | 1d 21h 35m | BTBBTTTB | +3.5% |
| 16 | 2d 12h | TBTT | 2d 12h | TTT | -/- |
| 17 | 1d 16h | TTB | 1d 16h | TTB (identical trip) | -/- |
| 18 | n/a | | 3d 5h 50m | BTTTBB | |
| 19 | 1d 9h | BBBT | 1d 10h 28m | BTT | +4.5% |
| 20 | 2d 13h | TTTBBB | 2d 3h | TTTBB | -20% |
| 21 | 2d 19h | BBBT | 2d 19h | BBBWBTT | -/- |
| 22 | 3d 7h | BTTBBB | 3d 6h 33m | BTBWBB | -/- |
| 23 | n/a | | 16h 38m | TB | |
| 24 | n/a | | 2d 15h 31m | BBTTT | |
| 25 | n/a | | 2d 13h | BBTTTTB | |

Table 2: results for 25 random trips – ground transport only

1) B = bus T = train P = plane W = walk X = taxi

Google Transit query times: 370-800 msec, average 540 msec

BBtime query times: 890-980 msec, average 920 msec. All trips except No. 15 return identical results at 500 msec query time.

| Trip # | air+ground | | ground only | |
|---|---|---|---|---|
| 1 | 7h 36m | TBWPXB | 28h 11 min | BTTWBW |
| 2 | 52m | BBW | 55m | BBW |
| 3 | 7h 58m | BTPWBW | 44h 50m | BWBBBWBW |
| 4 | 7h 9m | BWPXB | 31h | BBTTTB |
| 5 | 7h 59m | BXTWBW | 8h 4m | BBTWBW |
| 6 | 8h 2m | BTPPXBW | 13h 41m | BWTXBW |
| 7 | 8h 7m | BBBBBW | 8h | BBBBBW |
| 8 | 3h 51m | BTTXB | 4h 30m | BTTWBW |
| 9 | 6h 45m | BBWPB | 34h 45m | WBTTB |
| 10 | 20h 54m | BBBPBBB | 3d 21h 23 m | BTTBBB |
| 11 | 6h 48m | BXPTWB | 20h | WBTBB |
| 12 | 10h 6m | BTWPXB | 2d 1h 26m | BBWBTWB |
| 13 | 9h 49m | BTPPBWBW | 2d 6h | BWTTTWBW |
| 14 | 6h 8m | BXPWB | 6h 50m | WBTWBW |
| 15 | 8h | BTXPXB | 1d 21h 35m | BTBBTTTB |
| 16 | 18h 42m | TXPTWT | 2d 12h | TTT |
| 17 | 15h 48m | TWBPXB | 1d 16h | TTB |
| 18 | 29h 50m | BWBWPPXB | 3d 5h 50m | BTTTBB |
| 19 | 13h 33m | BBPPXT | 1d 10h 28m | BTT |
| 20 | 26h 38m | TXPPXBB | 2d 3h | TTTBB |
| 21 | 21h 23m | BXPPPXT | 2d 19h | BBBWBTT |
| 22 | 28h 58m | WBWBWPPXB | 3d 6h 33m | BTBWBB |
| 23 | 16h 38m | TB | 16h 38m | TB |
| 24 | 13h 10m | BXPPPX | 2d 15h 31m | BBTTT |
| 25 | 16 h 18m | PXPPPXTB | 2d 13h | BBTTTTB |

Table 3: comparison of ground-only and combined ground-air transport

1) B = bus T = train P = plane W = walk X = taxi

# Appendix B

Example trip results

Trip 6

From:  Park ave & newington sb, Baltimore, MD
       39.313856,-76.629285
To:    Lakeshore Dr at Elkwood Ave, Asheville, NC
       35.629776,-82.574842

summary:  8 hours 2 min  776.3 Km  6 stops

1. 20151213.936 Park ave & newington sb, Baltimore  39.3138559,-76.6292849
          bus  005 Mondawmin - Cedonia Maryland Transit  15 min  3.83 Km
   20151213.951 Eutaw st & saratoga st sb, Baltimore  39.2928469,-76.6210889

          Lexington st & light rail stat sb  39.2916769,-76.6197429 within 170 m
2. 20151213.1001 continue with   10 min  transfer time
           train  Light rail Maryland Transit  34 min  14.1 Km
   20151213.1035 Bwi airport lt rail sb  39.1812339,-76.6683639

          Baltimore / Washington BWI  39.1797740,-76.6688489 within 160 m
3. 20151213.1150 continue with  1 hour 12 min  transfer time
           plane AA2002 Baltimore to Charlotte  American Airlines  1 hour 36 min  580.6 Km
   20151213.1326 Charlotte CLT, NC  35.2208719,-80.9438889

4. 20151213.1429 continue with  1 hour transfer time
           plane AA4923 Charlotte to Asheville  American Airlines   51 min  146.5 Km
   20151213.1520 Asheville AVL,NC  35.4349970,-82.5381909

5. 20151213.1605 continue with   45 min  transfer time
           taxi     29 min  20.7 Km # (direct 17.6 Km)
   20151213.1634 Asheland Ave at Patton Ave  35.5940619,-82.5562220

6. 20151213.1701 continue with   27 min  transfer time
           bus  N1 North 1 Asheville Redefines Transit   23 min  9.5 Km
   20151213.1724 Elkwood Ave at Merrimon Ave  35.6371520,-82.5750019

7. 20151213.1724 continue
           walk     14 min  990 m   (direct 820 m)
   20151213.1738 Lakeshore Dr at Elkwood Ave, Woodfin, NC  35.6297760,-82.5748420

evaluated 35.73 K alternatives with 5.82 M departure times in 467 milliseconds

Trip 13

From:   Olive @ partridge sb, St Louis, MO
        38.667968,-90.319624
To:     NW Macleay & Lomita Terrace, Portland, OR
        45.526313,-122.706988

summary:  9 hours 49 min  3042.3 Km 7 stops next in  24 hours

1. 20151214.1425 Olive @ partridge sb  38.6679679,-90.3196240
        bus  91 Olive Metro St. Louis    8 min  3.15 Km
   20151214.1433 Des peres @ delmar loop metrolink eb  38.6552959,-90.2949519

        Delmar metrolink station  38.6557239,-90.2945909 within 50 m
2. 20151214.1439 continue with  6 min  transfer time
        train  MLR MetroLink Red Line Metro St. Louis   17 min  12.4 Km
   20151214.1456 Lambert main trml metrolink station  38.7412190,-90.3645999

        St Louis STL   38.7401779,-90.3640199 within 120 m
3. 20151214.1613 continue with  1 hour 16 min  transfer time
        plane DL2581 St Louis to Minneapolis  Delta Air Lines  1 hour 32 min  722.5 Km
   20151214.1745 Minneapolis MSP 44.8830910,-93.2107419

4. 20151214.1844 continue with   57 min  transfer time
        plane DL1411 Minneapolis to Portland  Delta Air Lines  3 hours 58 min  2289.1 Km
   20151214.2242 Portland PDX, OR 45.5888950,-122.5928650

5. 20151214.2330 continue with  48 min  transfer time
        bus  Downtown Express Blue Star Bus    5 min  10.5 Km
   20151214.2335 Marriott-Naito Pkwy. 45.5126720,-122.6756449

6. 20151214.2336 continue
        walk     19 min  1.29 Km   (direct 1.10 Km)
   20151214.2355 W Burnside & Burnside Bridge  45.5232489,-122.6710069

7. 20151215.2 continue with  7 min  transfer time
        bus  20 Burnside/Stark TriMet   9 min  2.97 Km
   20151215.11 W Burnside & NW Uptown Terrace  45.5242729,-122.7031579

8. 20151215.11 continue
        walk     3 min  260 m   (direct 210 m)
   20151215.14 NW Macleay & Lomita Terrace  45.5263129,-122.7069880

evaluated 60.16 K alternatives with 9.10 departure M times in 464 milliseconds